\documentclass[twoside,fleqn]{article}
\usepackage{espcrc2}
\usepackage{latexsym}

\newcommand{\be}{\begin{equation}}
\newcommand{\ee}{\end{equation}}
\newcommand{\<}{\langle}
\renewcommand{\>}{\rangle}
\newcommand{\reff}[1]{(\ref{#1})}

\title{Lattice QCD and chiral mesons }

\author{Sergio Caracciolo\address{Scuola Normale Superiore
    and INFN, Sezione di Pisa, I-56100 Pisa, ITALIA}, 
Fabrizio Palumbo\address{INFN -- Laboratori Nazionali di Frascati,
    P.~O.~Box 13, I-00044 Frascati, ITALIA }, and Roberto
    Scimia\address{Dip. di Fisica dell'Universit\`a  and INFN, Sezione di
    Perugia,  Via A. Pascoli,
    I-06100 Perugia, ITALIA} 
 }

\begin{document}

\begin{abstract}
The standard QCD action is improved by the addition of irrelevant
operators built  
with chiral composites. An effective Lagrangian is derived in terms of
auxiliary fields,  
which has the form of the phenomenological chiral Lagrangians. 
Our improved QCD action appears promising for numerical
simulations as the pion physics is explicitely accounted for by the
auxiliary fields.
\end{abstract}

\maketitle

QCD in the high energy limit is efficiently described by the
perturbative expansion in the gauge coupling 
constant. For the low energy properties instead, we do not have 
an equally satisfactory theory. 

At the phenomenological level a description of the low energy physics
related to the 
symmetry breaking of the chiral invariance is obtained by using the
phenomenological chiral Lagrangians~\cite{Wein,GL}, but a direct derivation
from the basic theory of quarks and gluons is still lacking.

At a more fundamental level the lattice formulation has allowed us to
investigate the quark confinement~\cite{Wilson} and the  spontaneous
breaking of chiral invariance, but due to the complexities related to
the definition of 
chiral fermions~\cite{Wilson,NN}, in numerical simulations the
chiral limit is achieved only through a fine-tuning procedure. Even
though these 
difficulties can be overcome, we remain unable to unify the treatment of
high energy and low energy properties.
 
For these reasons we have developed an approach based on the use of quark
composites as fundamental variables. The idea behind it is that a
significant part of the binding of the 
hadrons can be accounted for in this way, so that the ``residual
interaction" is sufficiently weak for a perturbative treatment.

The quark-composites approach is in principle fairly general, since it 
allows us to treat all the hadrons composite of quarks, but
for technical reasons the  composites with the quantum numbers of 
mesons and  barions are treated in a 
different way. The nucleonic composites, for instance, naturally satisfy
the Berezin integration rules and we 
derived the substitution rules which allow us to replace polynomials
of the quark fields by appropriate 
polynomials of these composites~\cite{DeFr} in the partition
function. The mesonic composites instead, due to the complexity of the 
integral over even elements of a Grassmann algebra, are 
replaced by auxiliary fields~\cite{Cara} by means of the
Stratonovich-Hubbard transformation~\cite{Stra}. Even though our
effective action is not 
renormalizable by power counting, due to the renormalizability 
of QCD  only a finite
number of free parameters can be generated by the counterterms because
of the BRS identities.   

We assume the modified partition function 
\begin{equation}
Z = \int [dV]  [d\overline{\lambda} d\lambda]\; \exp[- S_{YM} -S_{q} -S_C], 
\end{equation}
where $S_{YM}$ is the Yang-Mills action, $S_q$ is the action of the
quark  field  
and $S_C$ {\it is a four fermions irrelevant operator} which provides
{\em the kinetic terms } 
for the quark composites with the quantum numbers of the chiral
mesons.  $\lambda$ is the quark field while the
gluon  
field is  associated to the link variables $V$. Differentials
in square brackets  
are understood to be the product of the differentials  over the
lattice sites and  
the internal indices. All the fields live in an euclidian lattice of
spacing $a$.   
We introduce the following notation for the  sum over the lattice 
\be
(f,g)  =   a^4 \sum_x f(x) g(x).
\ee
In this notation the quark action is
\be
S_Q= (\overline{\lambda}, Q \lambda) .
\ee
We do the Wilson choice for  the quark wave operator
\be
Q  =   Q_D - Q_W,\,\,\,Q_D = \gamma_\mu  \, \overline{\nabla}_{\mu},\,\,\,
Q_W =   a \, {r \over 2} \, \Box. 
\ee
The symmetric derivative $ \overline{\nabla}_{\mu}$ and the Laplacian
$ \Box$ are  
covariant and are defined in terms of the right/left derivatives
\be
( \nabla^{\pm}_{\mu})_{x \, y}  =   \pm {1\over a} \left( \delta_{x
    \pm \hat{\mu},\,y}  V_{ \pm \mu} (x)
  -  \, \delta_{x \, y} \right) 
\ee
according to
\be
\overline{\nabla}_{\mu}  = { 1 \over 2} \left( \nabla^{+}_{\mu} +
  \nabla^{-}_{\mu} \right) , \,\,\,
\Box  =  \sum_\mu  \nabla^{+}_{\mu} \,  \nabla^{-}_{\mu}
\ee
with standard conventions for the link variables. 

The chiral composites are the pions and the sigma
\be 
  \vec{\hat{\pi}}  =  i\,a^{2}\,k_{\pi}\,\overline{\lambda} \gamma_5
  \vec{\tau}  \lambda,\;\;\;
  \hat{\sigma}  = a^{2}\, k_{\pi}\,\overline{\lambda} \lambda.    
\ee
$\gamma_5$ is assumed hermitian, the $\vec{\tau}$'s are the Pauli
matrices and a factor of dimension 
(length)${}^2$, necessary to give the composites the dimension of a
scalar, has been written in the form $a^2 k_{\pi}$ for convenience (see below).

Since for massless quarks the QCD action is chirally invariant, the
action of the chiral mesons must be, 
apart from a linear breaking term, $O(4)$ invariant. It must then have the form
\be
S_C=  { 1\over 4}\< (\hat{\Sigma}^\dagger,C \hat{\Sigma}) \> -
{ 1\over 4}\< (\hat{\Sigma}^\dagger + \hat{\Sigma}) m_q \>,\label{actionC2} 
\ee
where
\be
\hat{\Sigma} = \hat{\sigma} + i \,\vec{\tau} \cdot\vec{\hat{\pi}} \gamma_5 ,
\,\,\,\< A \> = \hbox{tr}^{isospin} A.
\ee
There is a long history of 4-fermions interactions and their relation to
QCD, starting from the Nambu-Jona-Lasinio~\cite{NJL} and  Gross-Neveu~\cite{GN}
models, until  the so-called chirally extended QCD or $\chi$QCD (see also for
example~\cite{Kogut:1997}).

Our demand of irrelevance of $S_C$, needed to maintain the renormalization
properties of QCD, and heuristic considerations, based on experience
with simple, solvable models, 
lead~\cite{Cara} to the following form of the wave operator of the 
chiral composites 
\be
 C = -{\rho^4 \over a^4} \; {1 \over - \Box + \rho^2 / a^2} \;, 
\ee
where $\rho$ is a dimensionless parameter.  
The irrelevance by power counting of $S_C$ requires that in the
continuum limit $\rho$ 
do not to vanish and $k_{\pi}$, as well as the product $k_{\pi} \rho$,
do not diverge.  
Under these conditions, as a consequence of the explicit dependence
on the lattice spacing of $C$ and the composites,  operators of dimension
higher than 4 are accompanied by the appropriate powers of the cut-off. 
All these parameters are obviously subject to renormalization.

We replace the chiral composites by the auxiliary fields 
\be
\Sigma = \Sigma_0 - i \gamma_5 \vec{\tau} \cdot \vec{\Sigma}  
\ee
 by means of the 
Stratonovich-Hubbard transformation~\cite{Stra}. Ignoring, as we will
systematically do in the sequel, 
field independent factors, the partition function can be written 
\begin{eqnarray} 
   Z  & =  &\int [dV]
   \left[{d\Sigma\over\sqrt{2 \pi}}\right]   \exp\left[-S_{YM} -  S_0
   \right]   \nonumber \\
& &  \int  [d \bar{\lambda} d\lambda]  \exp\left[ \left(\overline{\lambda},(D -
   Q)\lambda\right)   \right] \nonumber \\ 
 & = &
 \int [dV]  \left[{d{\Sigma}\over\sqrt{2 \pi}}\right]   
   \exp\left[-S_{YM} -  \tilde{S}  \right]    \label{mass} 
\end{eqnarray}
with
\be 
\tilde{S} = S_0 -  \mbox{Tr} \ln ( D - Q )
\ee
``Tr'' is the trace over both space and internal degrees
of freedom and we introduced  the functions of the auxiliary fields
\begin{eqnarray}
S_0  &=& -  {
   1\over 4} \rho^4 \< (\Sigma^\dagger, \left( a^4 \, C \right)^{-1} \Sigma) 
\>  ,  \label{schi} \\
D &=& k_\pi \left[ \rho^2 \Sigma - m_q\right].
\end{eqnarray}

The pion field can be introduced by the transformation
\begin{eqnarray}
\Sigma &= & R\, \left[ { 1 - \gamma_5 \over 2 } \, {U} +  { 1 +
    \gamma_5 \over 2 } \,{U}^{\dagger} \right]\nonumber\\
 R^2 &=& \Sigma_0^2 + \vec{\Sigma}^2 \nonumber \\
{U} &=&  \exp \left({i\over f_\pi} \vec{\tau}\cdot\vec{\pi}
    \right) \in SU(2). \label{tras} 
\end{eqnarray}
The volume element
\be
\left[{d{\Sigma}\over\sqrt{2 \pi}}\right] = \left[{dR \over \sqrt{2
      \pi}}\right ] [dU]\,\exp \sum_x 3 \ln R 
\ee
provides the Haar measure $[dU]$ over the group.
The effective action takes the form
\begin{eqnarray}
\tilde{S} &=& \sum_\mu {1\over 4} \<(\nabla_{\mu}^+ (R \,{U}^\dagger),
 \nabla_{\mu}^+ (R \,{U}) ) \> 
\nonumber\\
& & + { \rho^2 \over 2 a^2} (R,R)   - \mbox{Tr } \ln\, \left( D - Q \right). 
\label{effect}
 \end{eqnarray}
If the radial field $R$ acquires a nonvanishing expactation value 
$\overline{R}$, we set
\be
\overline{R} = f_\pi
\ee
so that the first term of $\tilde{S}$ is the kinetic part of the chiral
action while the radial field should not be dynamical because of its divergent 
mass (in the continuum limit). It can then be shown that the fermionic
determinant can be expanded in powers of the derivatives of $U$, as
appropriate to a Goldstone field, and of the explicit chiral symmetry 
breaking terms. 

If we could naively forget about the Wilson term in the
lattice action of the fermions, it would be very easy to evaluate 
\be
\overline{R}=  \sqrt{{\Omega} \over a \rho} = f_{\pi},
\ee
where $\Omega= 24$ is the number of quark components.
If  we neglect the fluctuations of $R$  and we put 
\be
B= { 1\over 2 a^2 f_{\pi}},
 \ee
the first term of $\tilde{S}$ can be identified with the leading term of the
chiral models 
\be
{\cal L}_2 =  {1\over 4}  f_{\pi}^2  \left[\< \nabla_{\mu} {U}^\dagger
\nabla^{\mu} {U} \>    
  - 2 m_q \, B   \<{U}  +  {U}^\dagger \> \right] 
\ee
and
\begin{eqnarray}
m_\pi^2 &= &  2 m_q B \\
k_\pi \<0| \overline{\lambda}\lambda |0\> &=& - 2 f_\pi^2 B
\end{eqnarray}
while the fermionic determinant has an (hopping) expansion in inverse powers of
$f_\pi$ and $k_\pi$  which provides corrections to
${\cal L}_2 $. Furthermore the quarks are perturbatively confined in this 
vacuum because their effective mass
\be
M_q = m_q - k_\pi \rho^2  \overline{R} =m_q - k_\pi \rho^2  f_\pi
\ee
is proportional to the inverse of the expansion parameters.

But since, as it is well known, in the absence of the Wilson term the
lattice action is describing 16 fermionic
species which cancel the abelian anomaly, let us go back to the full
theory in the presence of the term $Q_W$. 
Let us set $U=U^\dagger=1$ and $R=\overline{R}$ and study the
classical potential
\be
{\cal L} \left[\overline{R}\right] = {1\over 2} {\rho^2\over a^2}
\overline{R}^2 - {1\over 2 a^4 N^4} \hbox{ Tr } \ln P
\ee
where $a^4N^4$ is the lattice volume  
$$
P = M_q^2 - 2 M_q Q_W + Q_W^2 - Q_D^2 - [Q_D,Q_W]
$$
and the effective quark mass $M_q$ has been given above.

${\cal L}[\overline{R}]$ is still a function of the 
fluctuating gauge fields. The variation of the partition function
with respect to $\overline{R}$ yields the stationarity
equations
\be
 { \rho^2 \over a^2} \overline{R} = 
  { k_\pi \rho^2\over a^4 N^4 }   \, \<\<  \hbox{Tr }
 P^{-1} \left( Q_W - M_q \right) \>\>
\ee
where $\<\< \cdot \>\>$ is the functional average over the Yang Mills
measure.  Whenever 
 a nonvanishing solution to these  equations exists in the limit  of
vanishing of the explicit breaking terms, we
have {\em a spontaneous breaking of the chiral symmetry in QCD}.

Such solution must satisfy the condition on the effective quark mass
$M_Q$
\be
\lim_{a \rightarrow 0} { a M_q  \over  r}= 0. \label{anoma}
\ee
This is the standard condition, necessary also with our effective
action~\cite{CP}, for the  chiral anomaly to be correctly reproduced. 

The pion mass turns out to be
\be
m_{\pi}^2 = {k_\pi \rho^2 \overline{R}\over f_\pi^2} { 1 \over a^4 N^4} \<\<
 \mbox{Tr }  P^{-1} \left( Q_W - m_q \right) \>\>
\ee
and it is entirely due to the explicit breakings of the chiral invariance,
namely the quark mass term and the Wilson term. It is convenient to
introduce a subtracted mass which makes explicit the chiral limit
\be
m_q = m + \delta m
\ee
with
\be
  \delta m \<\< \mbox{Tr }   P^{-1} \>\> = \<\< \mbox{Tr }  P^{-1} Q_W  \>\>,
\ee
so that
\be
m_{\pi}^2 = - \, m { k_\pi \rho^2 \overline{R} \over f_\pi^2 }
{ 1 \over a^4 N^4} \<\< \mbox{Tr }  P^{-1} \>\>.
\ee
Since the quark condensate in the present case is given by
\be
\<0| \overline{\lambda} \lambda  |0\> =   \, { 1 \over a^4 N^4} \<\<
 \mbox{Tr }  P^{-1} \left(  k_{\pi} \rho^2 \overline{R}
     - m \right)  \>\>, \label{cond}
\ee
we get the Gell-Mann-Oakes-Renner relation~\cite{GMOR}
\be
m_{\pi}^2= -{ 1 \over f_{\pi}^2 } m  \<0| \overline{\lambda} \lambda  |0\>.
\ee
Eq.~\reff{cond} shows  a contribution to the effective quark mass 
proportional to the quark-condensate,  that has therefore the same
origin of the 
nonperturbative contribution to the quark mass  first studied by
Politzer~\cite{Politzer} (see also~\cite{Pascual}).

In conclusion, with our choice of the irrelevant 4-fermion $S_C$ we
get the desired link between QCD and the chiral lagrangians, provided
there exists a nontrivial solution for the expectation value of the
radial field $R$ which is related to the spontaneous symmetry breaking
of chiral invariance. It is indeed easy to show that the fermionic determinant
can be expanded in powers of the derivatives of the auxiliary fields and
the explicit chiral symmetry breaking terms. To complete the derivation of
the chiral models from QCD it remains to show that these latter will generate
only interactions of the chiral theories. This requires an investigation of the
the additive renormalizations which are induced by the Wilson term $Q_W$ 
through the appropriate Ward identities. It could be of interest, also in this
context, the use of Ginsparg-Wilson lattice fermions~\cite{GW} in order to
preserve as much as possible the explicit chiral invariance.
\par From a numerical point of view, the  evaluation of the fermionic
determinant should be faster in 
presence of the auxiliary fields, as they already provide the
propagation of the pions. Some support to the above can be found
in~\cite{Brower:1995}.


%

\end{document}